\documentclass[a4paper]{jpconf}
\usepackage{graphicx}
\begin{document}
\title{Effect of mass asymmetry on the mass dependence of balance energy.}

\author{Supriya Goyal}

\address{Department of Physics, Panjab University, Chandigarh 160 014, India}

\ead{ashuphysics@gmail.com}

\begin{abstract}
We demonstrate the role of the mass asymmetry on the balance
energy ($E_{bal}$) by studying asymmetric reactions throughout the
periodic table and over entire colliding geometry. Our results,
which are almost independent of the system size and as well as of
the colliding geometries indicate a sizeable effect of the
asymmetry of the reaction on the balance energy.
\end{abstract}

\section{Introduction}
Reaction dynamics has always captured a central place in nuclear
physics mainly due to the wider domain of physics it caters. Right
from the low energy (where fusion-fission takes place) to the
ultra-relativistic energies (where sub-nucleonic degrees of
freedom are dominant) a large number of new phenomena has been
observed/predicted. The reaction dynamics in heavy-ion collisions
at intermediate energies has been used extensively during the last
three decades to understand the nature of nuclear matter at
extreme conditions of temperature and density and the nature of
nuclear equation of state (EoS) \cite{1,2,3,3a,4,4a,4b}. The
phenomena that are mainly observed in this energy range are mainly
multifragmentation, collective transverse flow, particle
production etc.
\par
In the search of nuclear equation of state as well as of nuclear
interactions and forces, collective flow has been found to be of
immense importance \cite{1,2,3,3a,4,4a,4b}. Among collective flow,
transverse in-plane flow enjoys special status. During the last
two decades, much emphasis has been put on the study of collective
flow \cite{1,2,3,3a,4,4a,4b}. Lots of experiments have been
performed and number of theoretical attempts have also been
employed to explain and understand these observations. As reported
by \cite{5,5a} for the first time and later on by many others,
collective flow is negative at low incident energies whereas it is
positive at a reasonable higher incident energies. At a particular
incident energy, however, a transition occurs. This transition
energy is also known as energy of vanishing flow or balance energy
($E_{bal}$). This balance energy ($E_{bal}$) has been subjected to
intensive theoretical calculations using variety of equations of
state as well as nucleon-nucleon cross-sections
\cite{1,2,3,3a,4,4a,4b,5,5a}. This also includes the
mass-dependence of $E_{bal}$ which have been reproduced
successfully by various theoretical models \cite{1,3,3a,4,4a,4b}.
\par
Interestingly, most of these studies take symmetric or nearly
symmetric reactions into account. Recently, FOPI group studied the
flow for the asymmetric reaction of $^{40}Ca+^{197}Au$
\cite{6,6a}. They noted that the flow in the asymmetric collisions
is a key observable for investigating the reaction dynamics. In
contrast to symmetric collisions, where center of mass is one of
the nucleus, this quantity is not known {\it a priori} in
asymmetric nuclei experimentally. Later on, FOPI conducted
experiment on $^{58}Ni$ and $^{208}Pb$ \cite{6,6a}. In other class
of studies, the total mass of the system was kept fixed as 96
units whereas charge was varied \cite{ru}. Theoretically, recently
Kaur and Kumar \cite{7} conducted a complete study of the
multifragmentation by varying the asymmetry of the colliding
nuclei. Asymmetry parameter $\eta$ is defined as
($A_{T}-A_{p}$/$A_{T}+A_{P}$); where $A_{T}$ and $A_{P}$ are the
masses of the target and projectile, respectively. All these
attempts point towards a need for the systematic study of the
disappearance of flow for asymmetric colliding nuclei. Further as
noted, asymmetry of a reaction plays dramatic role in heavy-ion
collisions \cite{8,8a,8b,8c}. This happens because excitation
energy in symmetric colliding nuclei leads to larger compression
while asymmetric reactions lack the compression since large part
of the excitation energy is in the form of thermal energy.
\par
Note that some isolated studies with asymmetric nuclei are already
done in the literature where the reactions of $^{20}Ne+^{12}C$,
$^{20}Ne+^{27}Al$, $^{20}Ne+^{63}Cu$, $^{58}Ni+^{12}C$,
$^{64}Zn+^{27}Al$, $^{1}H+^{197}Au$, $^{12}C+^{197}Au$,
$^{197}Au+^{12}C$, $^{40}Ar+^{207}Pb$ etc are taken into account
\cite{6,6a,9,9a,9b,10,10a,10b,10c,10d,10e}. Interestingly, none of
the studies focus on the $E_{bal}$ for asymmetric colliding
nuclei. To address this, we here present a systematic study of the
$E_{bal}$ as a function of asymmetry of the colliding nuclei.
While total mass of the reactions remain fixed, the asymmetry is
varied by transferring the neutrons/protons from one nucleus to
other. The QMD model used for the present analysis is explained
briefly in the section 2. Results and discussion are presented in
section 3 followed by summary in section 4.

\section{\label{model}Description of the model}
The quantum molecular dynamics model (QMD) simulates the reaction
on an event by event basis \cite{2}. This is based on a molecular
dynamics picture where nucleons interact via two and three-body
interactions. The explicit two and three-body interactions
preserves the fluctuations and correlations which are important
for {\it N}-body phenomenon such as multifragmentation \cite{2}.
\par
In the QMD model, the (successfully) initialized nuclei are
boosted towards each other with proper center-of mass velocity
using relativistic kinematics. Here each nucleon ${\it \alpha}$ is
represented by a Gaussian wave packet with a width of $\sqrt{\it
L}$ centered around the mean position {\it $\vec{r}_{\alpha}$}(t)
and mean momentum {\it $\vec{p}_{\alpha}$}(t) \cite{2}:
\begin{equation}
\phi_{\alpha}(\vec{r},\vec{p},t)=\frac{1}{\left(2\pi
L\right)^{3/4}}e^{\left[-\left\{\vec{r}-\vec{r}_{\alpha}(t)\right\}^2/4L\right]}
e^{\left[i\vec{p}_{\alpha}(t)\cdot\vec{r}/\hbar\right]}.
\end{equation}
The Wigner distribution of a system with ${\it A_{T}+A_{P}}$
nucleons is given by
\begin{equation}
f(\vec{r},\vec{p},t)=\sum_{\alpha =
1}^{A_{T}+A_{P}}\frac{1}{\left(\pi \hbar\right)^{3}}
e^{\left[-\left\{\vec{r}-\vec{r}_{\alpha}(t)\right\}^2/2L\right]}
e^{\left[-\left\{\vec{p}-\vec{p}_{\alpha}(t)\right\}^2
2L/\hbar^{2}\right]^{'}},
\end{equation}
with L = 1.08 $fm^{2}$.
\par The center of each Gaussian (in the
coordinate and momentum space) is chosen by the Monte Carlo
procedure. The momentum of nucleons (in each nucleus) is chosen
between zero and local Fermi momentum
[$=\sqrt{2m_{\alpha}V_{\alpha}(\vec{r})};V_{\alpha}(\vec{r})$ is
the potential energy of nucleon $\alpha$ ]. Naturally, one has to
take care that the nuclei, thus generated, have right binding
energy and proper root mean square radii.
\par
The centroid of each wave packet is propagated using the classical
equation of motion \cite{2}:
\begin{equation}
\frac {d\vec{r}_{\alpha}}{dt} = \frac {dH}{d\vec{p}_{\alpha}},
\end{equation}
\begin{equation}
\frac {d\vec{p}_{\alpha}}{dt} = -\frac {dH}{d\vec{r}_{\alpha}},
\end{equation}
where the Hamiltonian is given by
\begin{equation}
H=\sum_{\alpha} \frac {\vec{p}_{\alpha}^{2}}{2m_{\alpha}} + V
^{tot}.
\end{equation}
 Our total interaction potential $V^{tot}$ reads as \cite{2}
\begin{equation}
V^{tot} = V^{Loc} + V^{Yuk} + V^{Coul} + V^{MDI},
\end{equation}
with
\begin{equation}
V^{Loc} = t_{1}\delta(\vec{r}_{\alpha}-\vec{r}_{\beta})+
t_{2}\delta(\vec{r}_{\alpha}-\vec{r}_{\beta})
\delta(\vec{r}_{\alpha}-\vec{r}_{\gamma}),
\end{equation}
\begin{equation}
V^{Yuk}=t_{3}e^{-|\vec{r}_{\alpha}-\vec{r}_{\beta}|/m}/\left(|\vec{r}_{\alpha}-\vec{r}_{\beta}|/m\right),
\end{equation}
with ${\it m}$ = 1.5 fm and $\it{t_{3}}$ = -6.66 MeV.
\par
The static (local) Skyrme interaction can further be parametrized
as:
\begin{equation}
U^{Loc}=\alpha\left(\frac{\rho}{\rho}_o\right)+
\beta\left(\frac{\rho}{\rho}_o\right)^{\gamma}.
\end{equation}
Here $\alpha, \beta$ and $\gamma$ are the parameters that define
equation of state. The momentum dependent interaction is obtained
by parameterizing the momentum dependence of the real part of the
optical potential. The final form of the potential reads as
\begin{equation}
U^{MDI}\approx t_{4}ln^{2}[t_{5}({\it\vec{p}_{\alpha}}-{\it
\vec{p}_{\beta}})^{2}+1]\delta({\it \vec{r}_{\alpha}}-{\it
\vec{r}_{\beta}}).
\end{equation}
Here ${\it t_{4}}$ = 1.57 MeV and ${\it t_{5}}$ = 5 $\times
10^{-4} MeV^{-2}$. A parameterized form of the local plus momentum
dependent interaction (MDI) potential (at zero temperature) is
given by
\begin{equation}
U=\alpha \left({\frac {\rho}{\rho_{0}}}\right) + \beta
\left({\frac {\rho}{\rho_{0}}}\right)^{\gamma}+ \delta
ln^{2}[\epsilon(\rho/\rho_{0})^{2/3}+1]\rho/\rho_{0}.
\end{equation}
The parameters $\alpha$, $\beta$, and $\gamma$ in above Eq. (11)
must be readjusted in the presence of momentum dependent
interactions so as to reproduce the ground state properties of the
nuclear matter. The set of parameters corresponding to different
equations of state can be found in Ref. \cite{2}.

\section{\label{results} Results and Discussion}
For the present study, we simulated various reactions for
1000-5000 events in the incident energy range between 90 and 500
MeV/nucleon. In particular, we simulated the reactions of
$^{17}_{8}O+^{23}_{11}Na$ ($\eta = 0.1$),
$^{14}_{7}N+^{26}_{12}Mg$ ($\eta = 0.3$),
$^{10}_{5}B+^{30}_{14}Si$ ($\eta = 0.5$), and
$^{6}_{3}Li+^{34}_{16}S$ ($\eta = 0.7$) for total mass ($A_{TOT}$)
= 40, $^{36}_{18}Ar+^{44}_{20}Ca$ ($\eta = 0.1$),
$^{28}_{14}Si+^{52}_{24}Cr$ ($\eta = 0.3$),
$^{20}_{10}Ne+^{60}_{28}Ni$ ($\eta = 0.5$), and
$^{10}_{5}B+^{70}_{32}Ge$ ($\eta = 0.7$) for total mass
($A_{TOT}$) = 80, $^{70}_{32}Ge+^{90}_{40}Zr$ ($\eta = 0.1$),
$^{54}_{26}Fe+^{106}_{48}Cd$ ($\eta = 0.3$),
$^{40}_{20}Ca+^{120}_{52}Te$ ($\eta = 0.5$), and
$^{24}_{12}Mg+^{136}_{58}Ce$ ($\eta = 0.7$) for total mass
($A_{TOT}$) = 160, and $^{108}_{48}Cd+^{132}_{56}Ba$ ($\eta =
0.1$), $^{84}_{38}Sr+^{156}_{66}Dy$ ($\eta = 0.3$),
$^{60}_{28}Ni+^{180}_{74}W$ ($\eta = 0.5$), and
$^{36}_{18}Ar+^{204}_{82}Pb$ ($\eta = 0.7$) for total mass
($A_{TOT}$) = 240. The present study is for peripheral collisions
(i.e. b/b$_{max}$ = 0.5). Note that in some cases, slight
variation can be seen for charges. The charges are chosen in a way
so that colliding nuclei are stable nuclides. A momentum dependent
soft equation of state with standard energy dependent cugnon
cross-section (labeled as SMD) is used for the present
calculations.
\begin{figure}[!t]
\centering
\includegraphics* [scale=0.7] {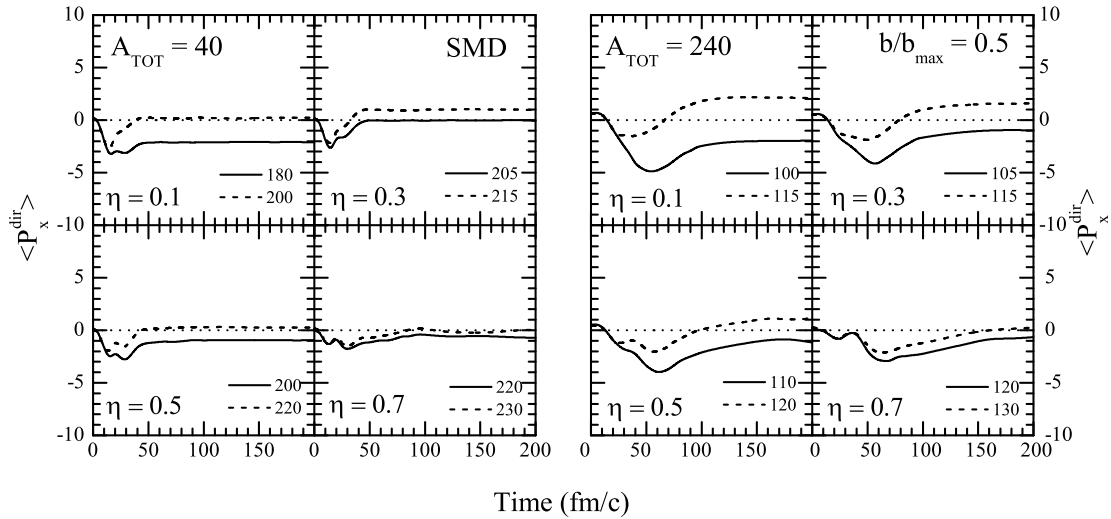}
\vskip -3.0cm \caption { The time evolution of the directed
transverse flow $<P^{dir}_{x}>$ as a function of time for
different mass asymmetries. Columns 1 and 2 from left are for
system mass ($A_{TOT}$) = 40, whereas columns 3 and 4 are for
system mass ($A_{TOT}$) = 240. Here we display the results for
peripheral collisions at different incident energies using SMD
equation of state.}
\end{figure}
\par
The balance energy ($E_{bal}$) is calculated using the {\it
directed transverse momentum $<P^{dir}_{x}>$}, which is defined
as:
\begin{equation}
\langle P_{x}^{dir}\rangle=\frac{1}{A}\sum_i {\rm
sign}\{Y(i)\}~{\bf{p}}_{x}(i),
\end{equation}
where $Y(i)$ and ${\bf{p}}_{x}(i)$ are the rapidity distribution
and transverse momentum of $i^{th}$ particle, respectively.
\par
In Fig. 1, we display the time evolution of the directed
transverse flow $<P^{dir}_{x}>$, for various asymmetric reactions
at different incident energies with system mass ($A_{TOT}$) = 40
and 240 units. From the figure, it is clear that transverse flow
is always negative during the initial phase of the reaction
irrespective of incident energy, system mass, and asymmetry of the
reaction. This shows that the interactions among nuclei are
attractive during initial phase of the reaction. These
interactions turn repulsive depending on the incident energy. The
transverse in-plane flow in lighter colliding nuclei for all
values of $\eta$ saturates earlier compared to heavy colliding
nuclei. A sharp transition from negative to positive flow is seen
for lighter nuclei at lower asymmetry. For the heavier nuclei at
all asymmetries; the transition is gradual. From the figure, one
can see that nearly symmetric reactions ($\eta$ = 0.1 and 0.3)
respond strongly to the change with incident energy compared to
highly asymmetric reactions. The cause behind is that with the
increase in the asymmetry of a reaction lesser binary collisions
take place resulting in lesser density and therefore, less
response occurs for variation in the collective flow. On the other
hand, for nearly symmetric reactions, the binary collisions
increases linearly with incident energy, therefore, huge
difference can be seen.
\begin{figure}[!t]
\centering \vskip -4.0cm
\includegraphics* [scale=0.42] {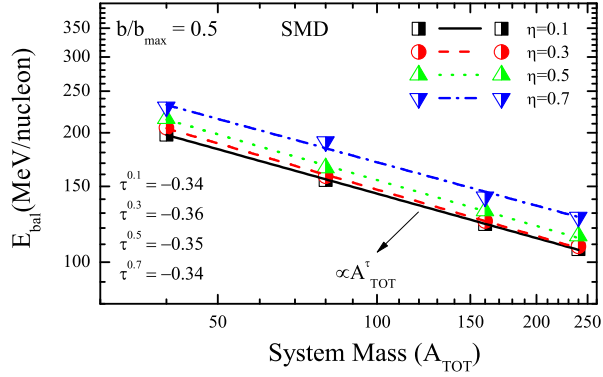}
\vskip -1.0cm \caption {(Color Online) The $E_{bal}$ as a function
of system mass ($A_{TOT}$) of reacting partners for SMD EoS and
for peripheral collisions. The results for different asymmetries
$\eta$ = 0.1, 0.3, 0.5, and 0.7 are represented, respectively, by
the half filled squares, circles, triangles and inverted
triangles. Lines are power law fit $\propto$ $A^\tau$.}
\end{figure}
\begin{figure}[!t]
\centering \vskip -1.0 cm
\includegraphics* [scale=0.42] {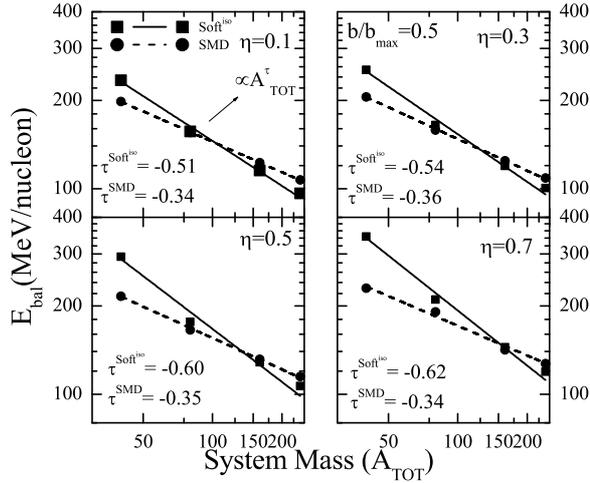}
\vskip -4.0 cm \caption { The $E_{bal}$ as a function of system
mass. Solid squares(circles) are for Soft$^{iso}$(SMD) EoS. The
top(bottom) left and right panels are for asymmetry parameter of
0.1(0.5) and 0.3(0.7) respectively. Lines are power law fit
$\propto$ $A^\tau_{TOT}$.}\label{fig3}
\end{figure}
\begin{figure}[!t]
\centering \vskip -1.0 cm
\includegraphics* [scale=0.42]{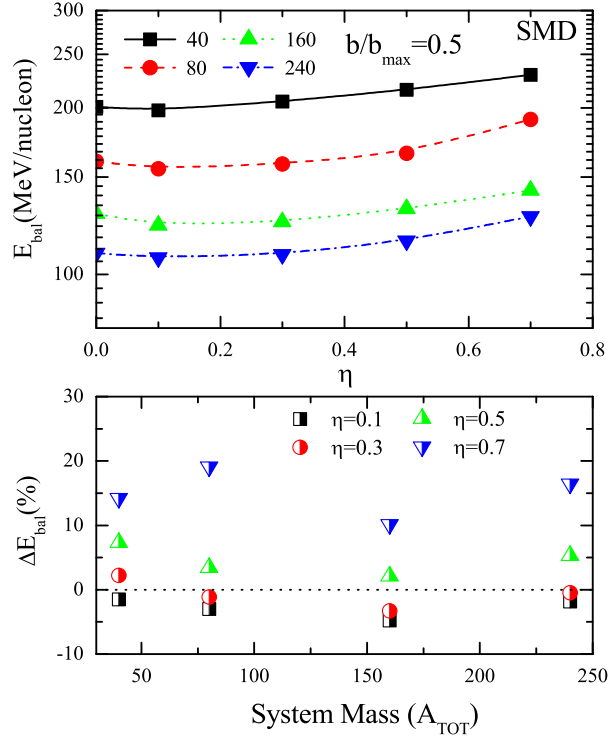}
\vskip -0.7cm \caption {(Color Online) $E_{bal}$ as a function of
asymmetry parameter $\eta$ is displayed in the upper panel,
whereas the percentage difference $\Delta$$E_{bal}(\%)$ as a
function of system mass of reacting partners is shown in the lower
panel. SMD EoS and an reduced impact parameter of 0.5 is used in
both cases. The results for the systems having total mass
$A_{TOT}$ of 40, 80, 160, and 240 are represented, respectively,
by the solid squares, circles, triangles and inverted triangles.
Lines are to guide the eye. The results of the percentage
difference for different asymmetries $\eta$ = 0.1, 0.3, 0.5, and
0.7 are represented, respectively, by the half filled squares,
circles, triangles and inverted triangles.}\label{fig4}
\end{figure}
\begin{figure}[!t]
\centering \vskip -1.0cm
\includegraphics* [scale=0.42]{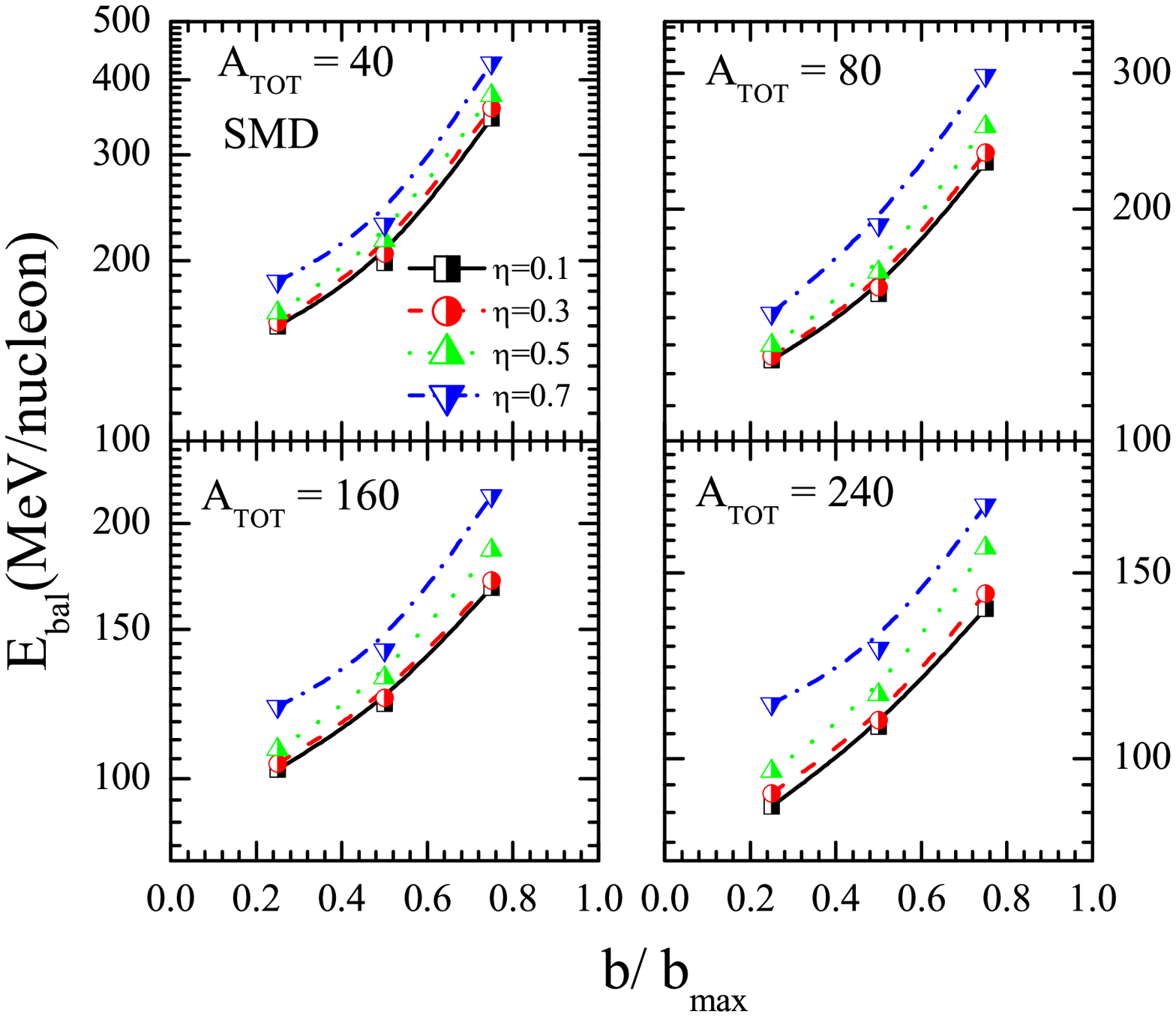}
\vskip -4.0cm\caption {(Color Online) The $E_{bal}$ as a function
of reduced impact parameter. The top(bottom) left and right panels
are for $A_{TOT}$ = 40(80) and 160(240) respectively. The results
for different asymmetries $\eta$ = 0.1, 0.3, 0.5, and 0.7 are
represented, respectively, by the half filled squares, circles,
triangles and inverted triangles. Lines are to guide the
eye.}\label{fig5}
\end{figure}
\par
It would be interesting to see how mass dependence character of
the $E_{bal}$ behaves at a fixed asymmetry. In Fig. 2, we display
the $E_{bal}$ verses combined system mass by keeping the asymmetry
fixed. We notice that in all the cases, a perfect power law
dependence (with power factor close to 1/3) can be seen for all
asymmetries right from 0.1 to 0.7. The values of power factor are
-0.34, -0.36, -0.35, and -0.34, respectively, for $\eta$ = 0.1,
0.3, 0.5, and 0.7. Strikingly, the results are similar to the one
as shown recently by Chugh and Puri \cite{chugh}. There it was
predicted that the role of MDI for symmetric systems is more
dominant at peripheral geometries and the value of power factor is
nearly equal to -0.35. From the figure, it is clear that all
points are lying on the line, indicating that asymmetry plays a
major role in the mass dependence. The variation in $E_{bal}$ for
$\eta$ varying between 0.1 and 0.7 is $\approx$ 32, 35, 19, and 19
for $A_{TOT}$ = 40, 80, 160, and 240, respectively. This explains
why in the earlier mass dependence studies \cite{1}, though
average behavior was a power law, individual $E_{bal}$ were quite
far from the average law \cite{1,3,3a}. There this happened
because no control was made for the asymmetry of a colliding pair.
Had all reactions been analyzed on a fixed asymmetry, one could
have obtain perfect match with average power law.
\par
The effect of MDI on the $E_{bal}$ is shown in Fig. 3, where we
display the $E_{bal}$ as a function of combined mass of the system
with Soft$^{iso}$ (Soft$^{iso}$ represents the soft equation of
state with isotropic standard energy dependent cugnon
cross-section) and SMD EoS for $\eta$ = 0.1, 0.3, 0.5, and 0.7.
Lines represent the power law fitting and different symbols are
explained in the caption of the figure. It is clear from the
figure that as $\eta$ increases, the suppression of $E_{bal}$ for
the lighter systems with MDI also increases while the increase in
$E_{bal}$ for heavier nuclei remains almost same. Since transverse
flow decreases with increase in asymmetry due to the decrease in
nn collisions and Coulomb repulsions, therefore, the effect of the
repulsive nature of MDI increases with increase in asymmetry
mainly for the lighter masses. With the increase of asymmetry of
the reaction, the role of MDI on the mass dependence of $E_{bal}$
is similar to its role for the symmetric reactions at
semi-peripheral and peripheral geometries.
\par
In Fig. 4, we display $E_{bal}$ as a function of $\eta$ for a
fixed mass equal to 40, 80, 160, and 240 units (top panel). In
agrement with all previous studies, $E_{bal}$ decreases with
increase in the mass of the system. This decrease has been
attributed to the increasing role of Coulomb forces in heavier
colliding nuclei. As discussed in earlier figures, a sizeable
influence can be seen towards $E_{bal}$ with variation of the
asymmetry of a reaction. For lighter masses, the effect of the
variation of $\eta$ can result about 35 MeV change in the
$E_{bal}$. As noted in absolute terms, lighter nuclei are more
affected compared to heavier ones. Overall, one sees that the
effect of the asymmetry of a reaction is not at all negligible. It
can have sizeable effect which goes as power law with power factor
close to 1/3. We also display the percentage difference
$\Delta$$E_{bal}$(\%) defined as $\Delta$$E_{bal}$(\%) =
(($E^{\eta\neq0}_{bal}$-$E^{\eta=0}_{bal}$)/$E^{\eta=0}_{bal}$)$\times$100
as function of system mass (bottom panel). Very interestingly, we
see that the effect of the asymmetry variation is almost uniform
throughout the periodic table. In other words, asymmetry of a
reaction can play significant role in $E_{bal}$ and the deviation
from the mean line can be eliminated if proper care is taken for
the asymmetry of colliding nuclei.
\par
In Fig. 5, we display the $E_{bal}$ as a function of the impact
parameter for different asymmetries. A well known trend i.e.
increase in the $E_{bal}$ with impact parameter can be seen for
all ranges of system mass. Further, as demonstrated by many
authors, the impact parameter variation is less effected in the
presence of MDI. The striking result is that the effect of mass
asymmetry variation is almost independent of the impact parameter.

\section{\label{summary} Summary}
Using the quantum molecular dynamics model, we presented a
detailed study of the balance energy with reference to mass
asymmetry. Almost independent of the system mass as well as impact
parameter, an uniform effect of the mass asymmetry can be seen at
the energy of vanishing flow. We find that for large asymmetries,
($\eta$ = 0.7), the effect of asymmetry can be 15\% with MDI and
in the absence it can be 40\%. This also explain the deviation in
the individual $E_{bal}$ from the mean values as reported earlier.

\section{Acknowledgments}
Author is thankful to Council of Scientific and Industrial
Research (CSIR) for providing the Junior Research Fellowship.

\smallskip
\end{document}